\documentclass{aa}
\usepackage{graphics}

\begin{document}

\thesaurus{03(11.01.2; 11.17.3; 12.03.1; 13.18.1; 13.20.1)}

\title{An evolutionary model for GHz Peaked Spectrum Sources\\ 
Predictions for high frequency surveys }

\author{G. De Zotti\inst{1} \and G.L. Granato\inst{1} \and L. Silva\inst{2} 
\and D. Maino\inst{2} \and L. Danese\inst{2} 
} 

\institute{Osservatorio Astronomico di Padova, Vicolo dell'Osservatorio 
5, I-35122 Padova, Italy \and SISSA, International School for Advanced Studies, 
Via Beirut 2-4, I-34014 Trieste, Italy }

\offprints{G. De Zotti (dezotti@pd.astro.it)}

\date{Received 19 July 1999 }
\titlerunning{Evolutionary model for GHz Peaked Spectrum sources} 
\authorrunning{G.~De~Zotti et al.} 

\maketitle

\begin{abstract}

We have explored, in the general framework of the ``young source'' scenario, 
evolutionary models for GHz Peaked Spectrum (GPS) galaxies and quasars 
which reproduce the observed counts, 
redshift and peak frequency distributions of currently available samples. 
Substantially different cosmological evolution properties are found for 
the two populations: the quasar luminosity function must evolve strongly 
up to $z\sim 1$, while the data on galaxies may be consistent with no 
evolution. The models show that GPS sources (mostly quasars) may comprise 
quite a significant fraction of bright ($S> 1\,$Jy) radio sources  
at $\nu \geq 30\,$GHz if the intrinsic distribution of peak frequencies 
extends up to $\sim 1000\,$GHz. In any case, however, their fraction decreases 
rapidly with decreasing flux and their contribution to small scale fluctuations 
in the frequency range covered by the forthcoming space missions MAP 
and Planck Surveyor is expected to be minor.

\keywords{Galaxies: active -- quasars: general -- 
Radio continuum: galaxies -- cosmic microwave background -- Submillimeter} 

\end{abstract}

\section{Introduction}

One of the major uncertainties affecting estimates of microwave fluctuations  
due to discrete extragalactic sources (Toffolatti et al. 1998; Sokasian 
et al. 1999; Gawiser et al. 1999) arises from the very poor knowledge  
of the abundance of radio sources with strongly rising spectra in the 
frequency range from a few tens to a few hundred GHz, where the most 
sensitive experiments to map primordial anisotropies of the Cosmic 
Microwave Background (CMB) are carried out. 

GHz Peaked Spectrum (GPS) radio sources [see O'Dea (1998) for a comprehensive 
review] selected at $\nu \leq 5\,$GHz have a fairly flat distribution of 
peak frequencies, suggesting the existence of an hitherto unknown 
population of sources peaking at mm wavelengths (O'Dea \& Baum 1997; 
Crawford et al. 1996; Lasenby 1996). Grainge \& Edge (1998) report the 
detection of 50 
GPS sources brighter than 50 mJy at 5 GHz and spectra still rising above 
10 GHz. The {\it observed} peak frequency of several of these sources was 
found to be $\geq 43\,$GHz; one has a redshift of 3.398 so that the 
emission peak is above 190 GHz in the rest frame.
  
It is thus possible that GPS sources significantly contaminate experiments 
aimed at obtaining very high accuracy (at the $\mu$K level), high resolution 
maps of the CMB, such as those expected from NASA's MAP and ESA's 
Planck Surveyor missions.

Unfortunately, the very limited information currently 
available makes it very difficult to carry out reliable predictions of 
the confusion noise due to these sources. On the other hand, physical models, 
although still schematic, may provide a useful guide for such predictions. 
An investigation of this kind may also be useful in connection with 
the on-going searches of GPS sources 
peaking at high frequencies (Guerra et al. 1998; Grainge \& Edge 1998; 
Cooray et al. 1998).

Two main scenarios have been proposed to explain the properties of GPS 
sources. One hypothesis is that their ages are similar to those of classical 
double radio sources and are kept compact for a large fraction of their 
lifetimes by interactions with dense gas in their environment  
(van Breugel et al. 1984; O'Dea et al. 1991; Carvalho 1998). 
Alternatively, GPS sources may correspond to the early stages of evolution 
of powerful radio sources (Phillips \& Mutel 1982; Carvalho 1985; 
Fanti et al. 1990, 1995; Readhead et al. 1996a,b; O'Dea \& Baum 1997). 
The latter possibility is currently favoured because the media surrounding 
radio sources appear to be insufficiently dense to inhibit the radio source 
growth for a large fraction of its lifetime (Fanti et al. 1995). Also, 
direct evidences of very short kinematic ages have been found by 
Owsianik \& Conway (1998), Owsianik et al. (1998, 1999) and 
Stanghellini (private communication).

In this paper we adopt the young, evolving sources scenario. We have 
elaborated (Sect. 2) on the analytic 
self-similar evolution model proposed by Begelman (1996, 1999) to 
account for the main 
observed population properties of GPS sources (Sect. 3). In Sect. 4 
we present predictions for counts of GPS sources at MAP and Planck Surveyor 
frequencies. Our main conclusions are summarized and discussed in Sect. 5.

\section{Outline of the adopted evolution model} 

We make the following assumptions:

\begin{enumerate}

\item The initial radio luminosity function (in units of Mpc$^{-3}\,
{\mathrm{d}\log L_i}^{-1}$) is described by a power law:
\begin{equation}
n(L_i) \propto \left({L_i\over L_\star}\right)^{-\beta}, 
\quad L_{i,{\rm min}} \leq L_{i} \leq L_{i,{\rm max}},
\end{equation}
where $L_i$ is the luminosity before absorption.

\item In the GPS phase, the properties of the sources are determined 
by the interaction of a compact, jet-driven, overpressured,  
non thermal radio lobe with a dense interstellar medium (Begelman 1996, 
1999; Bicknell et al. 1997); the timescale of the interaction is very 
short in comparison with the 
cosmological-expansion timescale, so that the luminosity 
evolution of individual sources occurs at constant $z$. As the radio 
lobe expands in the surrounding medium, the emitted radio power decreases 
with the source age, $\tau$, 
as $L_i \propto \tau^{-\eta}$, and its linear size 
$l$ increases as $l \propto \tau^{\epsilon}$. If the density of the surrounding 
medium scales with radius as $\rho_e \propto r^{-n}$, we have 
(Begelman 1996, 1999) $\eta = (n+4)/[4(5-n)]$ and $\epsilon = 3/(5-n)$. 
There is a clear anticorrelation between intrinsic turnover 
frequency, $\nu_p$, and linear size (O'Dea \& Baum 1997): $\nu_p \propto 
l^{-\delta}$, with $\delta \simeq 0.65$. It follows that $\nu_p$ scales 
with time as $\nu_p \propto \tau^{-\lambda}$, with $\lambda = \delta\epsilon$.

\item The spectra of GPS sources are described by:
\begin{equation}
L_\nu = L_p\times \left\{ \begin{array}{ll}
       (\nu/\nu_p)^{\alpha_a} & \mbox{if $\nu < \nu_p$} \\
       (\nu/\nu_p)^{-\alpha}  & \mbox{if $\nu > \nu_p$} 
       \end{array} 
       \right.
\end{equation}
with $\alpha_a=0.8$ and $\alpha=0.75$, the mean values found by Snellen et 
al. (1998b). 

\end{enumerate}

\noindent
As the radio lobe expands, the peak luminosity $L_p$ 
varies as the consequence of two factors: 
the decrease of the emitted radio power and the decrease of $\nu_p$. Hence:
\begin{equation}
L_p(\nu_p) = L_{p,i} \tau^p = L_{p,i} \left({\nu_p \over \nu_{p,i}}\right)^
{-p/\lambda} = L_{p,i} \left({\nu_p \over \nu_{p,i}}\right)^
{\eta/\lambda - \alpha}  \label{eq:Lp} 
\end{equation}
with $L_{p,i}(z) \propto L_{i}(z)$ and $p=-\eta+\alpha\lambda$. 

If the birth rate of GPS sources is constant on time scales much shorter than 
the cosmological-expansion timescale, the peak luminosity function per 
unit $\mathrm{d}\log L_p$ is: 
\begin{equation}
n(L_p) \propto L_p^{1/p}.
\end{equation}
Also, since, in this case, the number of sources of age $\tau$ within 
$\mathrm{d}\tau$ is simply proportional to $\mathrm{d}\tau$ and 
$\mathrm{d}\tau/\mathrm{d}\nu_p \propto  
(\nu_p/\nu_{p,i})^{-(1+1/\lambda)}$, the epoch dependent luminosity function 
at a given frequency $\nu$ (Mpc$^{-3}\,{\mathrm{d}\log L_\nu}^{-1}\,\mathrm{GHz}^{-1}$) 
writes:
\begin{eqnarray}
n(L_\nu, \nu_p,z) &=& n_0 (1+z)^3 \times \nonumber  \\ 
 & &  \times \left({L_{p,i}(L_\nu,\nu_p)\over 
 L_\star(z)}\right)^{-\beta}\,\left({\nu_p\over \nu_{p,i}}\right)^{-(1+
1/\lambda)}\ , 
\label{eq:FL}
\end{eqnarray}
where $L_\star(z)$ is the redshift-dependent normalization luminosity.  
We have assumed luminosity evolution and adopted a very simple parametrization 
for it:
\begin{equation}
L_\star(z) = L_0\times \left\{ \begin{array}{ll}
       (1+z)^k & \mbox{if $z < z_c$} \\
       (1+z_c)^k  & \mbox{if $z > z_c $} 
       \end{array} 
       \right. \label{eq:evol}
\end{equation}
The redshift $z_c$ at which luminosity evolution levels off is a model 
parameter. 
We have normalized monochromatic luminosities to $L_0=10^{32}\,\hbox{erg}\,
\hbox{s}^{-1}\,\hbox{Hz}^{-1}$. 

The luminosity function at a frequency $\nu_2$ is related to that at a 
frequency $\nu_1$ by:
\begin{equation}
n(L_{\nu_2}) = n(L_{\nu_1})\times \left\{ \begin{array}{ll}
       (\nu_2/\nu_1)^{-\alpha_a} & \mbox{if $\nu_1< \nu_2 < \nu_p$} \\
       (\nu_1/\nu_p)^{\alpha_a} (\nu_2/\nu_p)^{\alpha} & \mbox{if 
$\nu_1 < \nu_p < \nu_2$} \\
       (\nu_2/\nu_1)^{\alpha} & \mbox{if $\nu_p< \nu_1 < \nu_2$} 
       \end{array} 
       \right.
\end{equation}
The number counts per steradian 
of GPS sources brighter than $S_\nu$ at the frequency 
$\nu$, with an observed peak frequency $\max(\nu,\nu_{p,\rm{min}}) < \nu_{p,0} 
< \nu_{p,\rm{max}}$ are given by
\begin{eqnarray}
&N&(>S_\nu; \nu_{p,0} >\nu) =    
\int_0^{\min\left[z_f,z_m(S_\nu)\right]} \mathrm{d}z\,
{\mathrm{d}V\over \mathrm{d}z} \times \nonumber \\
&\times& \int_{\max[\nu_{p,\rm{min}}(1+z),\nu(1+z)]}^{\min[\nu_{p,\rm{max}}
 (1+z),\nu_{p,i}]} \mathrm{d}\nu_p    \times \nonumber \\               
&\times&  \int_{\log L_{\rm min}(S_\nu,z,\nu_p)}^{\log L_{\rm max}(z)} 
\mathrm{d}\log L_\nu\, n(L_\nu,\nu_p,z) \, ,
\end{eqnarray}
where $z_f$ is the redshift of formation of the first GPS sources (we have 
set $z_f=3.5$, the maximum observed redshift of a GPS source), $z_m$ is 
the maximum redshift at which sources can have a flux $\geq S_\nu$, 
$L_{\rm min}$ is the minimum luminosity of a source of given $z$ and 
$\nu_p$ yielding a flux $\geq S_\nu$, $\mathrm{d}V/\mathrm{d}z$ 
is the volume element within a solid angle $\omega$ 
in a Friedman-Robertson-Walker universe ($\Lambda =0$):
\begin{equation}
{\mathrm{d}V \over \mathrm{d}z} = {c\over H_0}\omega {d_L^2\over (1+z)^6(1+\Omega z)^{1/2}}
\end{equation}
\begin{equation}
d_L = {c\over H_0} z \left(1 + z {2-\Omega \over 2 +\Omega z + 
2(1+\Omega z)^{1/2}}\right)
\end{equation}
\begin{equation}
S_\nu = {L_\nu K(z) \over 4\pi d_L^2}
\end{equation}
\begin{equation}
K(z)=(1+z) {L_{\nu(1+z)}\over L_\nu} \, .
\end{equation}
Similarly, 
the number counts 
of GPS sources with an observed peak frequency 
$\nu_{p,\rm{min}} < \nu_{p,0} < \min(\nu,\nu_{p,\rm{max}})$
are given by
\begin{eqnarray}
&N&(>S_\nu; \nu_{p,0} < \nu) =    
\int_0^{\min\left[z_f,z_m(S_\nu)\right]} \mathrm{d}z\,
{\mathrm{d}V\over \mathrm{d}z} \times \nonumber \\
&\times& \int_{\nu_{p,\rm{min}}(1+z)}^{\min[\nu\,(1+z),\nu_{p,\rm{max}}(1+z),\nu_{p,i}]} 
\mathrm{d}\nu_p    \times \nonumber \\               
&\times&  \int_{\log L_{\rm min}(S_\nu,z,\nu_p)}^{\log L_{\rm max}(z)} 
\mathrm{d}\log L_\nu\, n(L_\nu,\nu_p,z) \, ,
\end{eqnarray}
Throughout this paper, we adopt $H_0=50\,\hbox{km}\,\hbox{s}^{-1}\,
\hbox{Mpc}^{-1}$ and $\Omega =1$.

The distribution of observed peak frequencies per unit $\mathrm{d}\nu_{p,0}$ 
in a flux limited sample, ${\cal N}(\nu_{p,0};>S_\nu)$, is given by: 
\begin{eqnarray}
&{\cal N}&(\nu_{p,0};>S_\nu) =    
 \int_0^{\min\left[z_f,z_m(S_\nu,\nu_{p,0})\right]} \mathrm{d}z\,
{\mathrm{d}V\over \mathrm{d}z} \times \nonumber \\
&\times& \int_{\log L_{\rm min}(S_\nu,z,\nu_{p,0}(1+z))}
^{\log L_{\rm max}(z,\nu_{p,0}(1+z))} 
\mathrm{d}\log L_\nu\, n[L_\nu,z,\nu_{p,0}(1+z)]
\end{eqnarray}

\section{Comparison with observations}

\subsection{Samples}

The available information mostly comes from the samples by Snellen et al. 
(1998b) and by Stanghellini et al. (1998). Marecki et al. (1999) have 
recently defined a relatively large sample, but still with a limited fraction 
of optical identifications and redshifts. A search of GPS sources peaking 
above 10 GHz has been carried out by A.C. Edge and co-workers; however only 
a very preliminary report has been published so far (Grainge \& Edge 1998). 

The sample by Snellen et al. (1998b) comprises 47 sources 
selected at 325 MHz 
with a limiting flux density of approximately 18 mJy; it was also required 
that $S(5{\rm GHz}) \geq 20\,$mJy. Whenever possible,  
an inverted spectrum between 325 and 609 
MHz was used as a criterion to select candidate GPS sources; otherwise, the 
selection was based on the 325--5000 MHz spectral index. 
The 325--609 MHz selection yielded 
14 sources over an area of 119 square degrees; this is probably a lower 
limit to the true areal density of GPS sources since the 5 GHz flux limit 
may introduce a bias against GPS sources with lower peak frequencies and 
steep spectra beyond the peak. The 325--5000 MHz selection yielded 33 
sources over 522 square degrees; the lower areal density, compared with 
the previous case, although not very statistically significant, may reflect 
a stronger bias against lower peak frequencies. The condition that 
spectra of    
candidate GPS sources can be fitted, within the observed frequency range, 
by a self-absorbed synchrotron spectrum, discriminates against sources with 
high values of $\nu_{p,0}$ ($> 8.4\,$GHz). Snellen et al. (1998a) obtained  
optical identifications and R and/or I magnitudes for 41 of them. 
Redshifts were measured by Snellen et al. (1999) for 19 of these. 
Since GPS galaxies show 
a tight apparent magnitude-redshift correlation (Snellen et al. 1996), 
redshift estimates can be derived from photometric data; 
best fit relationships are provided by Snellen et al. (1996).

Stanghellini et al. (1998) have defined a complete sample of 33 GPS sources 
brighter than 1 Jy at 5 GHz, over an area of approximately 24600 square 
degrees. All objects are optically identified and have magnitude estimates; 
redshifts are available for all but 5 of them. This sample is complementary 
to that of Snellen et al. (1998b) not only because of the selection at 
a much higher frequency and at brighter radio fluxes, but also because, with 
only four exceptions, the peak frequencies of sources are below the 
frequency of selection, while the opposite is true for the other sample.

The sample by Marecki et al. (1999) comprises 76 sources with 
5~GHz fluxes $>200\,$mJy over an area of about 7655 square degrees. 
The GPS classification, however, may be doubtful for 28 of these sources 
whose spectral shape is not well determined. 
As mentioned by the authors themselves, a further uncertainty arises from 
the variability, typical of flat spectrum sources, which may also lead to a 
misclassification of some sources as GPS's. Only 21 objects 
are identified (3 galaxies and 18 quasars) and have measured redshift.

The sample selected by Edge and co-workers is particularly well suited to 
investigate the abundance of sources peaking at very high frequencies which 
could be tricky contaminants of CMB anisotropy maps. Grainge 
\& Edge (1998) report the detection of 50 GPS sources brighter than 50 
mJy at 5 GHz and peak frequencies above 10 GHz over an area of about 2000 
square degrees; they point out that some GPS sources may have been 
missed because of various selection biasses and, therefore, their result 
should be regarded as a lower limit. No details on the redshift and 
peak frequency distributions are given.

\subsection{Constraints on model parameters}

GPS galaxies and quasars turn out to have different redshift, rest-frame 
peak frequency, linear size and radio morphology distributions 
(Stanghellini et al. 1996; Snellen et al. 1998a; 
Stanghellini et al. 1998). Thus, they must be treated as different 
populations. Hence, we have fitted separately the redshift and $\nu_{p,0}$ 
distributions of GPS galaxies and quasars in the samples by Stanghellini et 
al. (1998) and Snellen et al. (1998b). The counts reported by Grainge \& 
Edge (1998) were used as an additional constraint. The best fit values 
of the parameters were derived using the routine ``amoeba" (Press et al. 1992) 
exploiting the downhill simplex method in multidimensions.

It became clear quite soon that there is no need for cosmological evolution 
to account for the observed redshift distributions of GPS galaxies. 
Therefore, to minimize 
the number of free parameters, a no-evolution model was adopted [$L_\star(z) =
L_0$ in Eq.~(\ref{eq:FL})]. Only very 
few ``local'' GPS sources are known; therefore the local luminosity function 
cannot be determined directly. The power law luminosity function adopted here 
is fully characterized by four quantities: the normalization $n_0$, the slope 
$\beta$, the minimum and maximum luminosities. Additional model  
parameters are $\eta$ and $\lambda$. 

Table~\ref{tabfit} gives the best fit values of the parameters for 
both galaxies and 
quasars. The normalization of the luminosity function and the minimum and 
maximum luminosities refer to the frequency of 300~MHz.  

\begin{table}
\centering
\caption[]{Normalization constants and model parameters}
\label{tabfit}
\begin{tabular}{lcc}
 & &  \\ \hline \hline
 & & \\
 & Galaxies & Quasars \\
 & &  \\ \hline   
 & &  \\
$L_0\,(\hbox{erg}\,\hbox{s}^{-1}\,\hbox{Hz}^{-1})$ & \multicolumn{2}{c}
{$10^{32}$} \\
$\nu_{p,i}\,$(GHz) & \multicolumn{2}{c}{1000} \\
$z_f$ & \multicolumn{2}{c}{3.5} \\
$n_0\,[\hbox{Mpc}^{-3}\,(d\log L_\nu)^{-1}\,\hbox{GHz}^{-1}]$ & 
$5.0\times 10^{-8}$ & $6.1\times 10^{-13}$ \\
$L_{\rm min}(300\,\hbox{MHz})/L_0 $ & 6.7   &  0.36 \\
$L_{\rm max}(300\,\hbox{MHz})/L_0 $ & $2.9\times 10^{5}$ & $4.2\times 10^{5}$ \\
$\beta$     & 0.75  &  0.59 \\
$\eta $     & 10.7    & 9.6   \\
$\lambda$   & 8.7   & 3.9   \\
$z_c$       & --    &  1    \\
$k$         & --    & 12.2   \\ \hline \hline
\end{tabular}

\end{table}

\begin{figure}
\resizebox{\hsize}{!}{\includegraphics{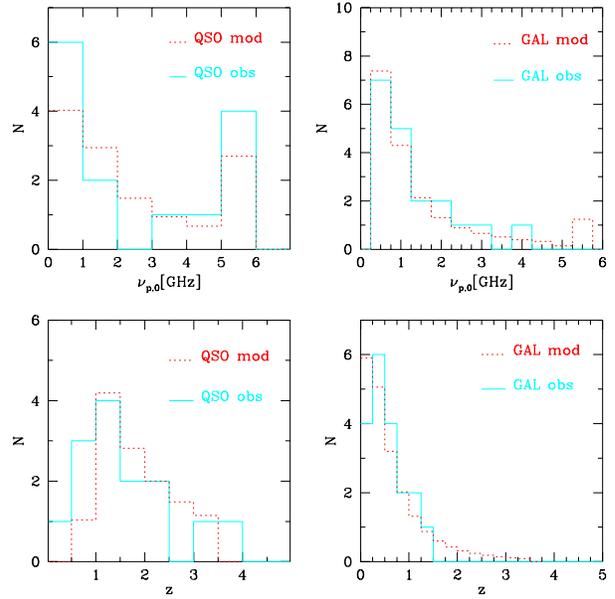}}
\caption{ 
Peak frequency and redshift distributions of quasars 
and galaxies in the sample by Stanghellini et al. (1998). The solid lines 
show the observed distributions. The dotted lines those implied by the model.}
\label{figstanghe}
\end{figure}

\begin{figure}
\resizebox{\hsize}{!}{\includegraphics{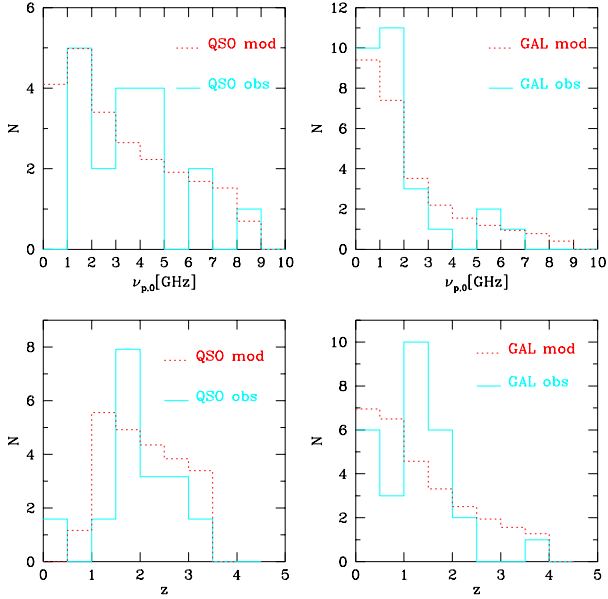}}
\caption{ 
Peak frequency and redshift distributions of quasars 
and galaxies in the sample by Snellen et al. (1998b). }
\label{figsnellen}
\end{figure}

Figs.~\ref{figstanghe} and \ref{figsnellen} compare the observed peak frequency and redshift distributions 
of GPS galaxies and quasars in the samples by Stanghellini et al. (1998) and 
Snellen et al. (1998b), respectively, with those yielded by the model. 

Redshifts of the 5 sources (four galaxies and 1 of uncertain identification, 
assumed to be a galaxy) in the sample by Stanghellini et al. (1998) 
lacking spectroscopic measurements were estimated using 
the magnitude-redshift relationships derived by Snellen et al. (1996).

We have assumed that the 19 objects in the sample by Snellen et al. (1998b), 
identified as star-like by Snellen et al. (1998a), are quasars. As suggested by 
Snellen et al. (1998a), we have adopted the faint source, rather than the 
quasar, as the identification of B1647$+$6225. Of those, 12 have measured 
redshifts (Snellen et al. 1999); their redshift distribution was taken as 
representative of the full sample (in practice, the observed number of 
quasars in each redshift bin was multiplied by 19/12). Only 6 of the 28 
``galaxies'' have a spectroscopic redshift; for the other alleged galaxies 
redshift estimates were derived from the magnitude-redshift relationships. 

The agreement between model and observed distributions is very satisfactory:  
the formal value of $\chi^2$ per degree of freedom is $\simeq 1$ both for 
galaxies and for quasars; a word of caution is in order, however,  
in view of the many uncertainties following from   
the various selection effects mentioned in the previous subsection (very 
difficult to quantify accurately) from photometric redshift estimates 
(especially for the sample by Snellen et al. 1998b) and from some doubtful  
identifications.

The model also yields 56 GPS sources (47 quasars and 9 galaxies) brighter than 
50$\,$mJy at $5\,$GHz and with $\nu_{p,0} > 10\,$GHz, over an area of 2000 
square degrees, to be compared with a 
lower limit of 50 sources (no further specifications given) reported by 
Grainge \& Edge (1998).

Assuming that the range of values of $\nu_0$ ($0.4\,\hbox{GHz} < \nu_0 < 
10\,\hbox{GHz}$) listed in Table~4 of 
Marecki et al. (1999) corresponds to the range of peak frequencies of 
GPS sources selected by these authors, the model yields 44 GPS sources 
(26 galaxies and 18 quasars) satisfying their selection criteria. As mentioned 
above, although the Marecki et al. (1999) sample comprises 76 sources, only 48 
of them have an unambiguous GPS classification.

We stated earlier that no evolution 
of GPS galaxies is required by the data. On the other hand, the derived 
evolutionary properties of GPS quasars are quite extreme, although the 
increase with redshift of the space density at any given luminosity is 
moderated by the flatness of the local luminosity function and by the 
relatively low value of the redshift at which the increase stops.  

The local luminosity functions of galaxies and quasars have almost 
identical slopes, significantly flatter than those of conventional 
radio sources (Dunlop \& Peacock 1990). The coincidence of the  
slopes of local luminosity functions of the two GPS populations may well 
be fortuitous. In the case of galaxies, the flat slope is required to 
reproduce redshift distributions 
extending to $z>3$ without resorting to cosmological evolution. In the case of 
quasars, a flat slope may be reminiscent of the effect of relativistic beaming 
on luminosity functions (Urry \& Shafer 1984; Urry \& Padovani 1991). 
Indications that quasar radio flux densities are moderately increased by 
Doppler boosting (while GPS galaxies are not) are discussed by O'Dea (1998).

An independent estimate of the local luminosity function of GPS sources was 
recently reported by Snellen \& Schilizzi (1999) who also found it to be 
flatter than that of steep spectrum radio sources. They argue that this 
is consistent with a scenario whereby GPS sources evolve to Compact Steep 
Spectrum (CSS) sources and to large-scale double radio sources, provided that  
the radio luminosity increases with time in the GPS phase and decreases 
in the subsequent phases.

Since only qualitative arguments are given by Snellen \& Schilizzi (1999) it 
is difficult to analyze quantitatively their model. We may mention, however, 
that we were unable to obtain a consistent fit to the data sets considered 
here with parameter values implying an increase of the radio power with 
increasing age of GPS sources. On the other hand, according to Eqs.~(2) and 
(3), the monochromatic radio luminosity, $L_\nu$, of GPS sources at frequencies 
$\nu < \nu_p$ varies as $L_\nu \propto \tau^p\nu_p^{-\alpha_a} \propto 
\tau^{p-\lambda \alpha_a}$. Inserting the values of parameters given in 
Table~1, it turns out that, in the case of galaxies (but not in the case of 
quasars), $L_\nu$ does {\it increase} with time ($\propto \tau^{2.8}$). 

The parameters $\eta$ and $\lambda$, characterizing the evolution with the 
source age of the emitted power and of the turnover frequency, are much 
larger than expected in the framework of the self-similar evolution model 
by Begelman (1996, 1999). The physical meaning of this result needs to be 
further investigated.

\section{High frequency counts of GPS sources}

\begin{figure}
\resizebox{\hsize}{!}{\includegraphics{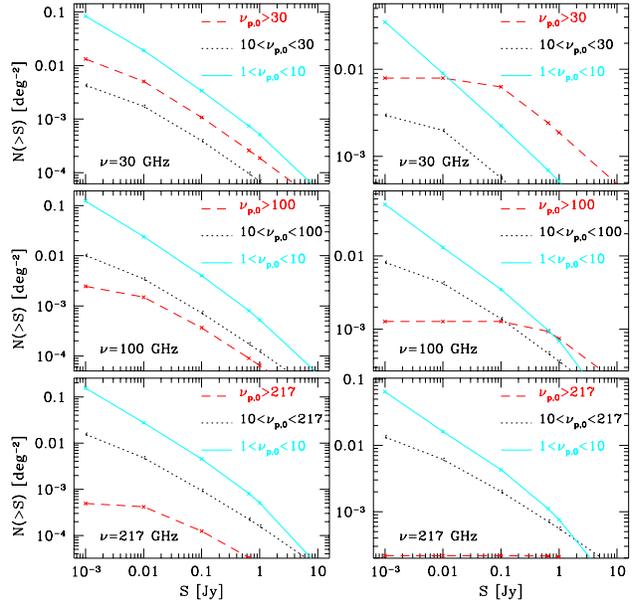}}
\caption{
Predicted counts of GPS galaxies (left-hand column) and quasars (right-hand 
column) at three Planck Surveyor frequencies for different 
ranges of observed $\nu_{p,0}\,$(GHz) and a maximum intrinsic value $\nu_{p,i}
=1000\,$GHz.} \label{gps1000} 
\end{figure}

\begin{figure}
\resizebox{\hsize}{!}{\includegraphics{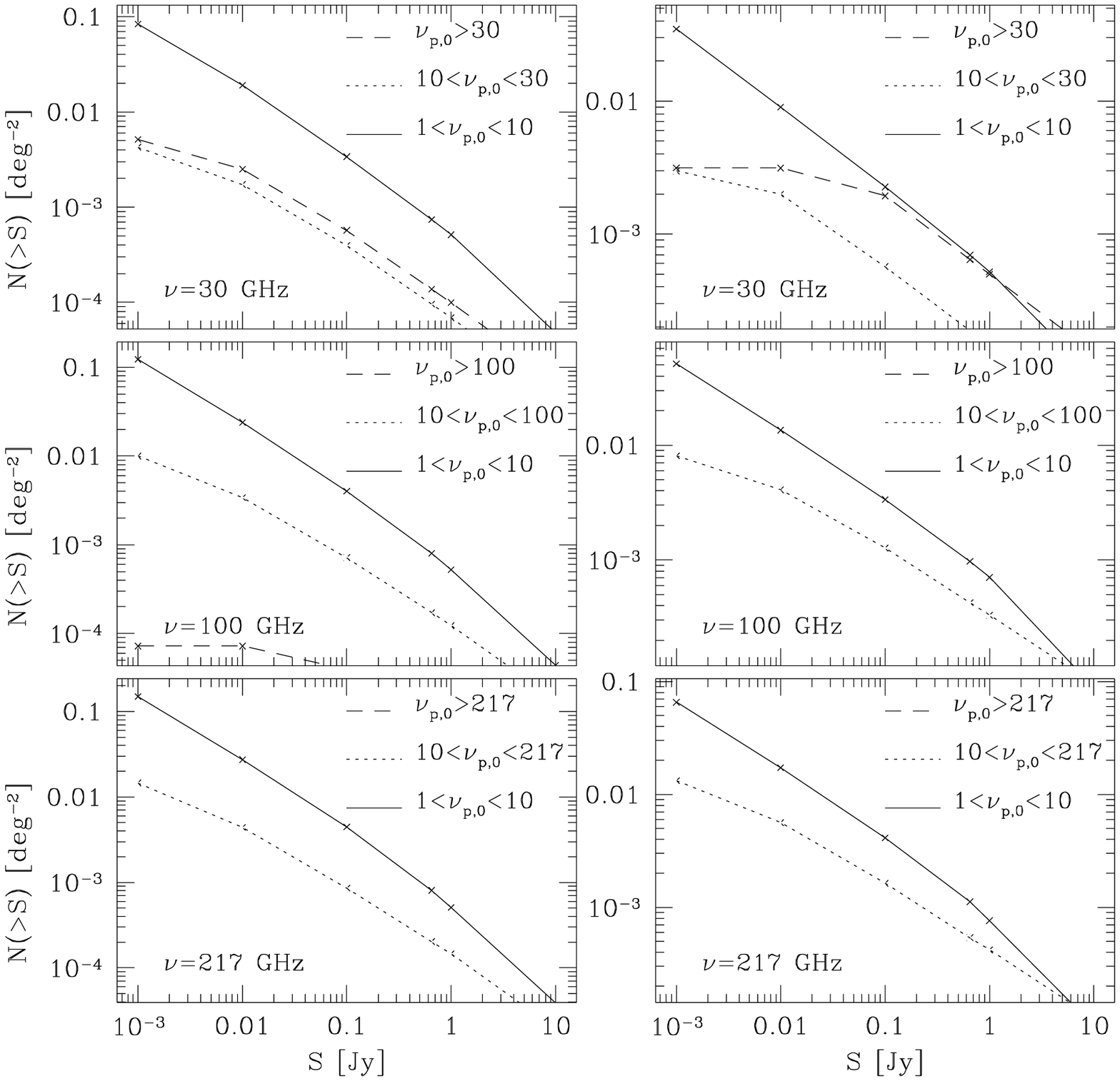}}
\caption{ Predicted counts of GPS galaxies (left-hand column) and 
quasars (right-hand column) at three Planck Surveyorfrequencies for different 
ranges of observed $\nu_{p,0}\,$(GHz) and a maximum intrinsic value $\nu_{p,i} 
=200\,$GHz.} \label{gps200}
\end{figure}

The high frequency counts of GPS sources further depend on the
maximum value, $\nu_{p,i}$, of the rest-frame peak frequency. We have 
considered two extreme cases. The results shown in Fig.~\ref{gps1000} 
for three Planck Surveyor frequencies   
correspond to a very large value, $\nu_{p,i}=1000\,$GHz. On the other hand, 
since there is direct evidence that values of $\nu_p \geq 190\,$GHz do exist 
(Grainge \& Edge 1998), we may take $200\,$GHz as a lower limit to 
$\nu_{p,i}$. The corresponding counts are shown in Fig.~\ref{gps200}. 

The results confirm the rough preliminary estimate by De Zotti et al. (1998): 
from several tens to a few hundreds of such sources could be detectable by  
Planck Surveyor instruments. 

Not surprisingly, given the high luminosities of GPS sources at low 
frequencies and their inverted spectra, sources with the highest peak 
frequencies (mostly quasars, whose peak frequency distribution is weighted 
towars higher values, as shown by Figs.~\ref{figstanghe} and \ref{figsnellen}),
show up at the brightest flux levels. If the rest-frame peak frequencies 
can reach very large values (up to $\simeq 1000\,$GHz), high-$\nu_{p,0}$ 
sources are expected to dominate the bright end of GPS source counts  
up to $\simeq 100\,$GHz. A comparison of the results shown in 
Fig.~\ref{gps1000} 
with the estimates by Toffolatti et al. (1998) and Sokasian et al. 
(1999) suggests that GPS sources may, in this case,  
correspond to a substantial fraction (30 to 70\%) of bright radio sources  
($> 1\,$Jy) at $\nu>30\,$GHz. On the other hand, since their model counts are  
rather flat, the fraction of GPS sources rapidly decreases with decreasing 
fluxes; at 0.1 Jy their fraction should drop to $\sim 5\%$, implying  
that they are a minor contributor to small scale 
fluctuations due to extragalactic sources.

The areal density of sources peaking at very high frequencies obviously 
decreases rapidly with decreasing $\nu_{p,i}$. For example, the $30\,$GHz 
counts of GPS quasars brighter than 1$\,$Jy 
with observed $\nu_{p,0}> 30\,$GHz decreases by about a 
factor of 4 as $\nu_{p,i}$ decreases from 1000 to 200$\,$GHz 
(see Fig.~\ref{gps200}), while the 
corresponding total counts of GPS sources decreases by about a factor of 2. 

Given their rather low 
areal density, only very large area surveys would allow the selection of 
significant samples of these sources. Planck Surveyor's all sky surveys are 
unique in this respect.

\section{Conclusions}

In the general framework of the ``young source'' scenario  
we have worked out exploratory models reproducing the observed counts, 
redshift and peak frequency distributions of currently available samples 
of GPS galaxies and quasars. The derived values for parameters characterizing 
the evolution with the source age of the emitted power and of the turnover 
frequency, however, are substantially larger than those usually quoted. 
Therefore, the physical interpretation of these results needs to be further 
investigated. 

With the due allowance for 
the many selection effects and uncertainties related to 
the identification of bona fide GPS sources, the agreement between model 
predictions and observational data appears to be satisfactory.

Substantially different evolution properties are required for GPS galaxy and 
quasar populations, while the derived local luminosity functions of the 
two populations have a similarly flat slope. This similarity may, however, 
be fortuitous, particularly if the radio emission of GPS quasars (but not 
of galaxies) is boosted, and the luminosity function flattened (Urry \& Shafer 
1984), by a relativistic Doppler effect, as suggested 
by some observational data (O'Dea 1998).

The model predicts that, if the peak frequency distribution extends to 
very large values ($\sim 1000\,$GHz in the rest frame), GPS sources (mostly 
quasars) may comprise 
quite a significant fraction (30 to 70\%) of bright ($S> 1\,$Jy) radio sources 
at $\nu \geq 30\,$GHz, according to estimates by Toffolatti et al. (1998) and 
Sokasian et al. (1999). If the maximum intrinsic peak frequency is $\simeq 
200\,$GHz [an intrinsic $\nu_p > 190\,$GHz has been reported by Grainge \&  
Edge (1998)] the number of GPS quasars brighter than 1 Jy at 30 GHz decreases 
by about a factor of 4 and the total areal density of GPS sources at the 
same flux limit decreases by a factor of about 2. 

In any case, the counts of GPS sources are expected to
be substantially flatter than those of other classes of extragalactic sources.
As shown by Figs.~3 and 4, the slopes of integral counts of both galaxies and
quasars never exceed 1 and are frequently substantially smaller. This
result may be surprising in the case of quasars, given the strong
luminosity evolution of their population. One should remember, however, that
as $z$ increases, the observational selection picks up higher and higher 
values of intrinsic peak frequencies and earlier and earlier
phases, when the luminosity is higher, of the evolution of individual
sources [cf. Eq.~(\ref{eq:Lp})]. The duration of early phases, however,
rapidly decreases with increasing luminosity; the effect is, in some sense,
equivalent to a negative density evolution which compensates for the positive
luminosity evolution of the quasar population. This ``negative evolution''
effect is much weaker in the case of galaxies because their redshift
distribution is weighted towards lower values and because the evolution
of individual sources is less rapid ($p=-\eta+\alpha\lambda \simeq -4.2$ for
galaxies while $p \simeq - 6.7$ for quasars).

The flatness of counts obviously implies that the fraction of GPS sources
quickly decreases with decreasing flux; as a consequence,
their contribution to small scale fluctuations in the frequency range 
covered by Planck Surveyor and MAP missions is
expected to be minor.

In closing, we stress once again the exploratory nature of this study. 
Several conclusions are still tentative and will need to be reassessed as 
more data will become available. 

\begin{acknowledgements}
We gratefully acknowledge very informative 
discussions with R. Fanti, C. Fanti and C. Stanghellini. A. Edge kindly 
provided some additional information on his sample. We wish to thank the 
referee for a careful reading of the manuscript and helpful comments. 
This work has been done in the 
framework of the Planck-LFI Consortium activities; it has been   
supported in part by ASI and MURST.
\end{acknowledgements}


\begin{thebibliography}{}

\bibitem[]{}
Begelman M.C., 1996, in {\it Cygnus A: Study of a Radio Galaxy}, ed. C. 
Carilli, D. Harris, Cambridge University Press, p. 209

\bibitem[]{}
Begelman M.C., 1999, in The Most Distant Radio Galaxies, proc. of 
the KNAW colloquium held in Amsterdam, 15-17th October 1997, 
eds. H.J.A. R\"ottgering, P.N. Best, M.D. Lehnert, KNAW, Amsterdam, p. 173

\bibitem[]{}
Bicknell G.V., Dopita M.A., O'Dea C.P., 1997, ApJ 485, 112

\bibitem[]{}
Carvalho J.C., 1985, MNRAS 215, 463

\bibitem[]{}
Carvalho J.C., 1998, A\&A 329, 845

\bibitem[]{}
Cooray A.R., Grego L., Holzapfel W.L., Joy M., Carlstrom J.E., 1998, 
AJ 115, 1388

\bibitem[]{}
Crawford T., Marr J., Partridge R.B., Strauss M.A., 1996, ApJ 460, 225

\bibitem[]{}
De Zotti G., Toffolatti L., Granato G.L., 1998, in 
{\it Fundamental Parameters in Cosmology},
Proc. XXXIII Rencontres de Moriond, eds. J. Tr\^an Thanh V\^an, 
Y. Giraud-H\'eraud, F. Bouchet, et al., Ed. Fronti\`eres, p. 143

\bibitem[]{}
Dunlop J.S., Peacock J.A., 1990, MNRAS 247, 19


\bibitem[]{}
Fanti C., Fanti R., Dallacasa D., et al., 1995, A\&A 302, 317

\bibitem[]{}
Fanti R., Fanti C., Schilizzi R.T., et al., 1990, A\&A 231, 333

\bibitem[]{}
Gawiser E., Jaffe A., Silk J., 1999, ApJ, submitted

\bibitem[]{}
Grainge K., Edge A., 1998, in {\it Fundamental Parameters in Cosmology}, 
Proc. XXXIII Rencontres de Moriond, eds. J. Tr\^an Thanh V\^an, 
Y. Giraud-H\'eraud, F. Bouchet, et al., 
Ed. Fronti\`eres, p. 151

\bibitem[]{}
Guerra E.J., Haarsma D.B., Partridge R.B., 1998, BAAS 193, 40.03
 
\bibitem[]{}
Lasenby A. N., 1996, in {\it Microwave Background Anisotropies}, eds. F.R.
Bouchet, R. Gispert and B. Guiderdoni, Ed. Fronti\`eres, p. 453

\bibitem[]{}
Marecki A., Falcke H., Niezgoda J., Garrington S.T., Patnaik A.R., 1999, 
A\&AS 135, 273

\bibitem[]{}
O'Dea C.P., 1998, PASP 110, 493

\bibitem[]{}
O'Dea C.P., Baum S.A., 1997, AJ 113, 148 

\bibitem[]{}
O'Dea C.P., Baum S.A., Stanghellini C., 1991, ApJ 380, 66

\bibitem[]{}
Owsianik I., Conway J.E., 1998, A\&A 337, 69

\bibitem[]{}
Owsianik I., Conway J.E., Polatidis A.G., 1998, A\&A 336, L37

\bibitem[]{}
Owsianik I., Conway J.E., Polatidis A.G., 1999, proc. 4th EVN/JIVE symp., 
New Astron. Rev., in press

\bibitem[]{}
Phillips R.B., Mutel R.L., 1982, A\&A 106, 21

\bibitem[]{}
Press W.H., Teukolsky S.A., Vetterling W. T., Flannery B. P., 1992, 
Numerical Recipes in Fortran - The Art of Scientific Programming, Second 
Edition, Cambridge University Press

\bibitem[]{}
Readhead A.C.S., Taylor G.B., Xu W., Pearson T.J., Wilkinson P.N., 
Polatidis A.G., 1996a, ApJ 460, 612

\bibitem[]{}
Readhead A.C.S., Taylor G.B., Pearson T.J., Wilkinson P.N., 1996b, ApJ 
460, 634

\bibitem[]{}
Snellen I.A.G., Bremer M.N., Schilizzi R.T., Miley G.K., van Ojik R., 1996,  
MNRAS 279, 1294

\bibitem[]{}
Snellen I.A.G., Schilizzi R.T., 1999, in Perpectives on Radio Astronomy: 
Scientific Imperatives at cm and mm Wavelengths, eds. M.P. van Haarlem, 
J.M. van der Hulst, NFRA, Dwingeloo 

\bibitem[]{}
Snellen I.A.G., Schilizzi R.T., Bremer M.N., et al., 1998a, MNRAS 
301, 985

\bibitem[]{}
Snellen I.A.G., Schilizzi R.T.,  de Bruyn A.G., et al., 1998b, A\&AS 131, 435

\bibitem[]{}
Snellen I.A.G., Schilizzi R.T., Bremer M.N., et al., 1999, MNRAS 307, 149

\bibitem[]{}
Sokasian A., Gawiser E., Smoot G.F., 1999, ApJ, submitted

\bibitem[]{}
Stanghellini C., Dallacasa D., O'Dea C.P., et al.,  1996, 
in proc. {\it 2nd workshop on GPS \& CSS Radio Sources}, eds. I. Snellen, 
R.T. Schilizzi, H.J.A. R\"ottgering, M.N. Bremer, p. 4

\bibitem[]{}
Stanghellini C., O'Dea C.P., Dallacasa D., et al., 1998, A\&AS 131, 303

\bibitem[]{}
Toffolatti L., Arg\"ueso-G\'omez F., De Zotti G., et al., 1998, MNRAS 297, 117 

\bibitem[]{}
Urry C.M., Padovani P., 1991, ApJ 371, 60

\bibitem[]{}
Urry C.M., Shafer R.A., 1984, ApJ 280, 569

\bibitem[]{}
van Breugel W., Miley G., Heckman T., 1984, AJ 89, 5

\end{thebibliography}
\end{document}